# Million-Degree Plasma Pervading the Extended Orion Nebula

Manuel Güdel[1,2,3,4], Kevin R. Briggs[1,5], Thierry Montmerle[4], Marc Audard[6,7], Luisa Rebull[8] and Stephen L. Skinner[9]

[1]Paul Scherrer Institut, Würenlingen and Villigen, CH-5232 Villigen PSI, Switzerland; [2]Max-Planck-Institute for Astronomy, Königstuhl 17, D-69117 Heidelberg, Germany; [3]Leiden Observatory, Leiden University, PO Box 9513, 2300 RA Leiden, The Netherlands; [4]Laboratoire d'Astrophysique de Grenoble, Université Joseph Fourier – CNRS, BP 53, 38041 Grenoble Cedex, France; [5]Institute of Astronomy, ETH Zentrum, CH-8092 Zürich, Switzerland; [6]ISDC, Chemin d'Ecogia 16, 1290 Versoix, Switzerland; [7]Geneva Observatory, University of Geneva, Ch. des Maillettes 51, 1290 Sauverny, Switzerland; [8]Spitzer Science Center, California Institute of Technology, Mail Code 220-6, Pasadena, CA 91125; [9]CASA, UCB 389, University of Colorado, Boulder, CO 80309-0389, USA.

**Most stars form as members of large associations within dense, very cold (10-100 K) molecular clouds. The nearby giant molecular cloud in Orion hosts several thousand stars of ages less than a few million years, many of which are located in or around the famous Orion Nebula, a prominent gas structure illuminated and ionized by a small group of massive stars (the Trapezium). We present X-ray observations obtained with the X-ray Multi-Mirror satellite XMM-Newton revealing that a hot plasma with a temperature of 1.7-2.1 million K pervades the**

**southwest extension of the nebula. The plasma, originating in the strong stellar winds from the Trapezium, flows into the adjacent interstellar medium. This X-ray outflow phenomenon must be widespread throughout our Galaxy.**

The Orion Nebula (M42, Fig. 1) together with its parent Orion Molecular Cloud is the prototypical massive star formation region *(1)* in which thousands of stars are forming from collapsing cloud cores *(2,3)*. Many forming stars are surrounded by molecular envelopes, or show circumstellar disks and bipolar jets *(4)*.

The Huygens region of the nebula is the bright region surrounding the massive, ionizing stars of the Trapezium group, the most important member being θ¹Ori C *(5)*. This region is a thin layer of ionized (HII) gas lying on the surface of the background molecular cloud, with the ionizing stars offset towards the observer. There is a foreground veil of neutral material that defines the central cavity. To the southwest of the Huygens region, the brightness drops rapidly as one views the much larger Extended Orion Nebula (EON henceforth; *(6)*). Ionized gas flows through the thin layer at about 10 km s$^{-1}$ and then into the southwest cavity *(5,7,8)*, where its density of about 30 cm$^{-3}$ *(9)* is much lower than the 10$^4$ cm$^{-3}$ in the Huygens region *(5)*. The geometry of the EON is not well understood, but it must be confined on the far side by the molecular cloud and seems to be laterally bounded by neutral gas and dust.

The Orion Nebula has been a favorite target of X-ray satellites for 30 years *(10-12)*, the Chandra X-ray Observatory having detected almost every single Orion member star as a vigorous X-ray source in a 17' x 17' field centered on the Trapezium *(12)*. In this paper, we describe deep X-ray observations with the X-Ray Multi-Mirror satellite

XMM-Newton *(13)* that afford much wider spatial coverage of the region. The half-degree diameter field we discuss here is centered to the southwest of the Trapezium, almost fully covering the EON cavity.

XMM-Newton's CCD cameras *(14,15)* observed several fields covering this region (Fig. 2A). The combined image shows the anticipated large assembly of X-ray bright stars, but also exhibits faint, extended structure in the EON, roughly consisting of a brighter patch in the north and a more extended area in the south. Overlaying the X-ray extended structure on an infrared image taken by the Spitzer Space Telescope (at 4.5 and 5.8μm, Fig. 2B), we see that the X-ray feature fills in the areas between the infrared structures formed by dense, warm dust, such as the shocks *(6)* in the west and bright lanes in the south of the cavity.

The spectrum of the extended emission (Fig. 3) shows a strong (unresolved) O VII line triplet at 0.55 keV, indicative of 1-2 MK plasma. Fitting the derived X-ray spectra with a hot plasma model reveals a dominant temperature of 1.7-2.1 MK *(16)*. The observed X-ray luminosity is proportional to the product of the square of the electron density, $n_e$, and the volume, V, the so-called the volume emission measure (EM = $n_e^2$V), for which the spectral fit found $(1.5\pm0.3) \times 10^{54}$ cm$^{-3}$ and $(1.9\pm0.3) \times 10^{54}$ cm$^{-3}$ for the northern and southern area, respectively. Emission from hotter plasma is also present in the spectrum but has been successfully modeled as residual counts from the (incompletely) excised point sources that show much harder emission than our extended structure *(16)*. The much harder stellar emission and the absence of a dense group of young stars in this region in the Spitzer images show that the soft X-rays cannot be unresolved stellar emission but are genuinely diffuse.

Chandra has not detected any diffuse X-ray emission in the central Huygens region *(17)*, most likely because the dense veil of neutral gas located in front of this region, but not of the EON, absorbs soft X-rays. The very low neutral gas column densities we measure in the spectra of the two regions of observed diffuse X-ray emission [$N_H = (4.1 \pm 0.7) \times 10^{20}$ cm$^{-2}$ for the northern region and $\leq 10^{20}$ cm$^{-2}$ at the 1$\sigma$ level for the southern region] support this view and conclusively demonstrate that the hot plasma is located in front of the dense molecular gas and therefore also the rather thin, optically visible ionization front on the surface of the molecular cloud.

The energy requirement to heat the large-scale X-ray emitting plasma is severe. The absorption corrected, intrinsic X-ray luminosity of the observed diffuse source is $L_X = 5.5 \times 10^{31}$ erg s$^{-1}$ in the 0.1-10 keV range (or $\approx 3.4 \times 10^{31}$ erg s$^{-1}$ pc$^{-2}$; additional, absorbed radiation may be present). Although massive molecular flows emanate from within the embedded cloud region, their velocities are typically <100 km s$^{-1}$ *(18)*, too slow to shock-heat the gas to the observed temperatures. The total power in fast microjets ejected by the numerous young stars falls short of the observed X-ray power by about two orders of magnitude *(4)*. The structure of the Orion Nebula, its young age [$\approx$3 Myr *(19)*] and the absence of radio shell structures *(9)* argue against a hot supernova bubble such as those observed in star-forming regions in the Large Magellanic Cloud *(20)*. The only efficient energy source is provided by the fast winds from the hot Trapezium stars. The wind from $\theta^1$ Ori C alone has a kinetic energy rate $L_w = \dot{M}V_w^2/2 \approx 7\times10^{35}$ erg s$^{-1}$ [$\dot{M} \approx 8 \times 10^{-7}$ M$_\odot$ yr$^{-1}$ being the mass-loss rate *(21)* and $V_w \approx 1650$ km s$^{-1}$ being the wind terminal velocity *(21)*], i.e., 4 orders of magnitude larger than the observed $L_X$; the high velocity wind can easily heat the observed plasma.

The ionization structure and dynamics of the Orion Nebula is very complex and not easily accessible to comprehensive modeling *(5)*. We can, however, gain insight by applying a simplified, classic interstellar bubble model *(22)*. This model of a massive star wind interacting with a surrounding uniform neutral gas yields four quantities: the radius of the shocked wind bubble $R_s$, its temperature and density $T_b$ and $n_b$, and its luminosity $L_b$, using as input parameters the density, $n_0$, of the gas surrounding the expanding wind bubble, the stellar age t, and the wind parameters $\dot{M}$ and $V_w$. Because of the weak power-law dependence on the input parameters, $T_b$ is robustly of order 1 MK and $n_b$ is of order 1 cm$^{-3}$ [with $n_0 = 10^4$ cm$^{-3}$ *(8)* and t = $10^6$ yr for $\theta^1$ Ori C, one finds $T_b$ = 2.4 MK and $n_b$ = 1.3 cm$^{-3}$]. Given the simplistic model, the computed bubble radius, $R_b \approx 3.9$ pc, is acceptably close to the observed EON cavity radius of $\approx 2$ pc (Figs. 1,4), especially because the plasma is not entirely confined (in contrast to the model) by the surrounding molecular material.

The cavity depth is likely to be at least 0.9 pc [the distance between the absorbing veil and the ionization front in the Trapezium region *(5)*] and no more than 4 pc (for a spherical volume filling the EON cavity). With the estimated projected areas of 0.24 pc$^2$ and 1.4 pc$^2$ for the northern and the southern diffuse X-ray regions, respectively, the observed X-ray emission measures suggest electron densities of $n_e \approx 0.2$-0.5 cm$^{-3}$ and 0.1-0.2 cm$^{-3}$, respectively. Because the density of the radio-emitting material is ≈100-300 times higher and its temperature of $(0.5-1) \times 10^4$ K is ≈200-400 times lower *(5,9)*, the two pressures, $p \approx 2n_e kT$, are comparable, i.e., pressure equilibrium approximately holds between the HII gas and the hot plasma if they are in contact. This suggests a leaking cavity in which plasma is continuously replenished. Therefore, the hot X-ray gas is likely to be channeled by the cooler, denser structures rather than disrupting them by expansion.

Extrapolating to the entire cavity (Fig. 4; cavity radius ≈ 2pc), the above densities suggest a total mass in the hot gas of of $0.07 M_\odot$ (for both geometric models). The mass loss rate of $\theta^1$ Ori C, $8 \times 10^{-7} M_\odot$ yr$^{-1}$ *(21)*, implies replenishment of this gas in only ≈$10^5$ yrs, much shorter than the radiative cooling time of 1.8-3.9 Myr derived from the density of the southern source. Because near-pressure equilibrium has been established, the hot gas must flow out of the cavity, in the form of an X-ray champagne flow *(7)*: in the case of our observed plasma, mass conservation for the wind of $\theta^1$ Ori C implies a plasma bulk velocity of a few 10 km s$^{-1}$, depending on the adopted geometry for the channel confined by the cavity. This outflow phenomenon may be common to all massive star-forming regions. The existence of wind-powered extended X-ray emission, theoretically predicted over 30 years ago *(22)*, was demonstrated only recently by Chandra observations of very massive star-forming regions like M17 and Rosette (spectral types O5 and earlier, or $M_* > 60 M_\odot$), the hot gas however showing harder emission *(17)*. Here, in contrast, we suggest that diffuse X-ray emission is commonly present in "classic", Orion-like HII regions that host fewer, less massive O stars with powerful winds.

The most likely outlet for the X-ray flow is the nearby Eridanus superbubble discovered by the Röntgen satellite (ROSAT) *(23)*. This huge, hot bubble, 20 degrees in diameter (≈140 pc), is thought to be the result of several supernova explosions having taken place in previous generations of Orion OB stars *(24)*. The wind-shocked gas leaking from the Orion Nebula would thus continuously replenish the Eridanus superbubble. Also, the Orion/Eridanus region has long been known to emit 1.809 MeV gamma-rays, corresponding to the decay of $^{26}$Al in less than a million years. Although the exact site remains uncertain, a production by the Orion massive stars, past or present, followed by funneling into the Eridanus superbubble, has been suggested *(25)*. Our X-ray observations provide a natural explanation for the existence of such a funnel.

Our Galaxy (and other star-forming galaxies) could thus maintain a network of X-ray bubbles and plasma flows, cooling over a few million years but continuously being replenished by shocked winds from a multitude of "modest" Orion-like star forming regions, gently leaking out the parent molecular clouds, in addition to being fed by discrete, but rare, supernova explosions.

References and Notes:

1. S. W. Stahler, F. Palla, *The Formation of Stars* (Wiley-VCH, Weinheim, 2005).
2. M. J. McCaughrean, J. R. Stauffer, *Astron. J.* **108**, 1382-1397 (1994).
3. L. A. Hillenbrand, L. W. Hartmann, *Astrophys. J.* **492**, 540-553 (1998).
4. J. Bally, C. R. O'Dell, M. J. McCaughrean, *Astron. J.* **119**, 2919-2959 (2000).

29. We thank David Malin for providing the optical image of the Orion Nebula taken with the UK Schmidt telescope and granting permission to use it, and Ravi Subrahmanyan and the American Astronomical Society for granting permission to reproduce the radio panel in Fig. 4D. This research is based on observations obtained with XMM-Newton, an ESA science mission with instruments and contributions directly funded by ESA member states and the USA (NASA). MA acknowledges support from a Swiss National Science Foundation Professorship (PP002-110504), and SS from NASA grant NNG05GE69G.

30. **Supporting Online Material**
    www.sciencemag.org
    Materials and Methods
    Table S1
    Figures S1–S3

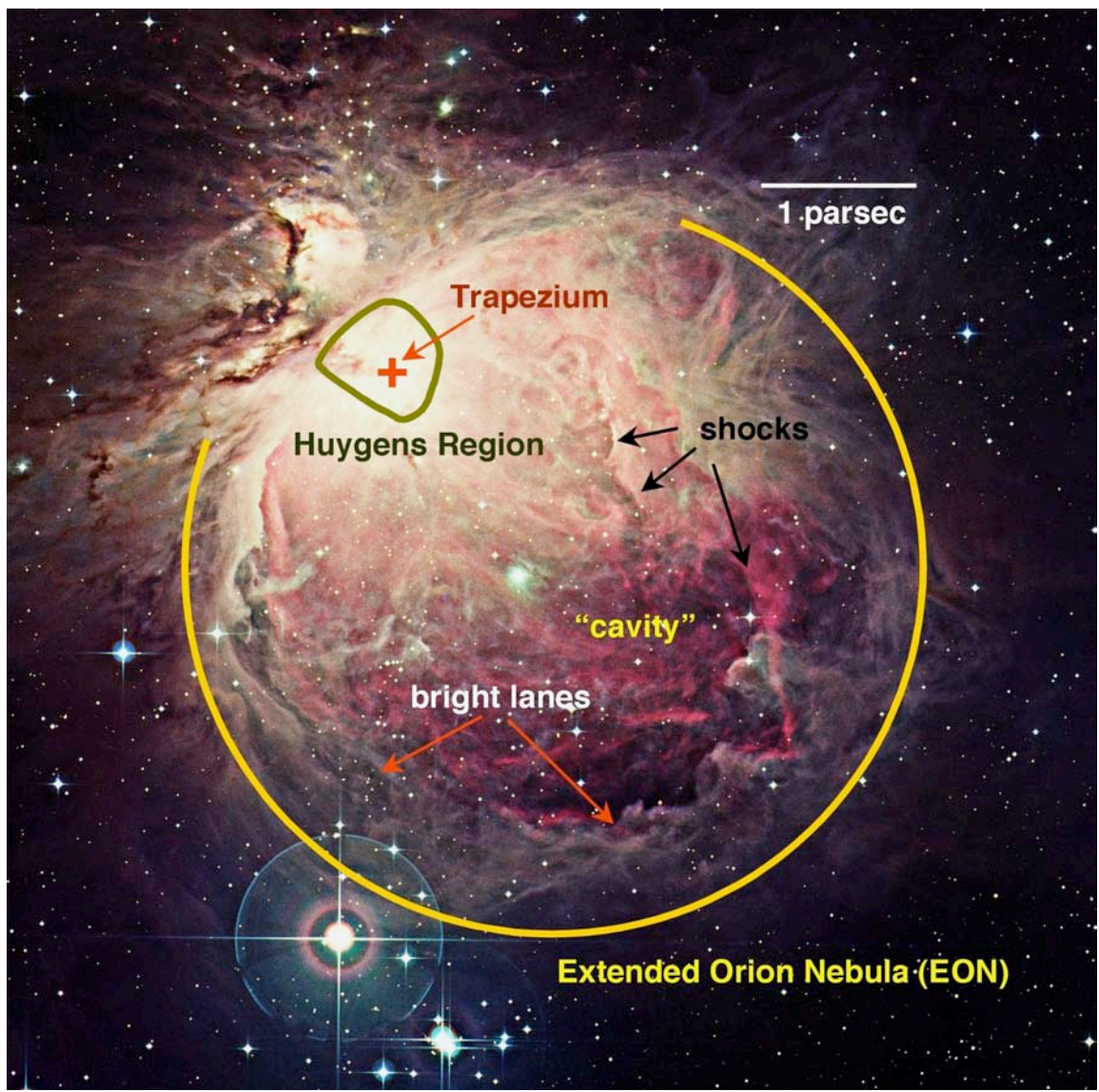

Figu re 1: The Orion Nebula in the optical light (UK Schmidt telescope; copyright Anglo-Australian Observatory/David Malin Images). Features described in the text are labelled. The bar in the upper right corner gives the length of one parsec ($3.1 \times 10^{18}$ cm). North is up and west is to the right.

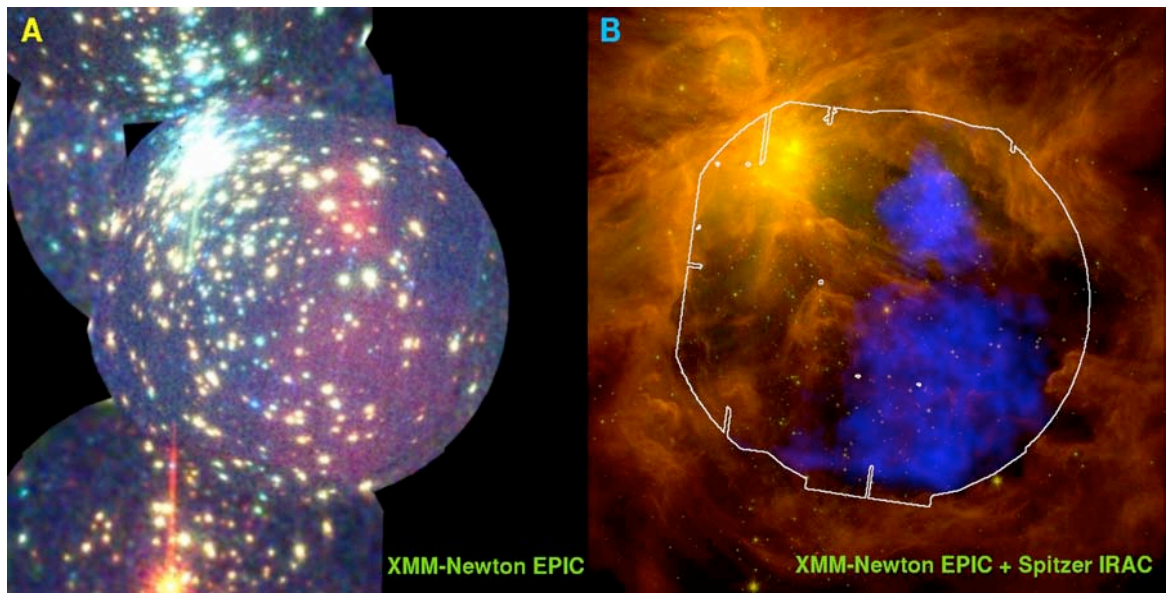

Figure 2: The Orion Nebula with its hot gas bubble. The X-ray image (A) is color-coded for photon energies in the 0.3-7.3 keV range (red to blue). The diameter of each of the near-circular fields is 30 arcminutes (3.5 pc), and the angular resolution is approximately 5 arcseconds. Panel (B) shows, on the same scale, the excess diffuse emission in the 0.3-1 keV band with respect to the hard band extracted from the longest observation *(16)* in blue, overlaid on a composite 4.5μm (green channel) and 5.8 μm (red channel) mid-infrared image from the Spitzer Space Telescope. X-ray point sources have been removed, and the residual image has been adaptively smoothed *(16)*. The intensity scale is logarithmically compressed. The white contour shows the detector field-of-view for this X-ray observation.

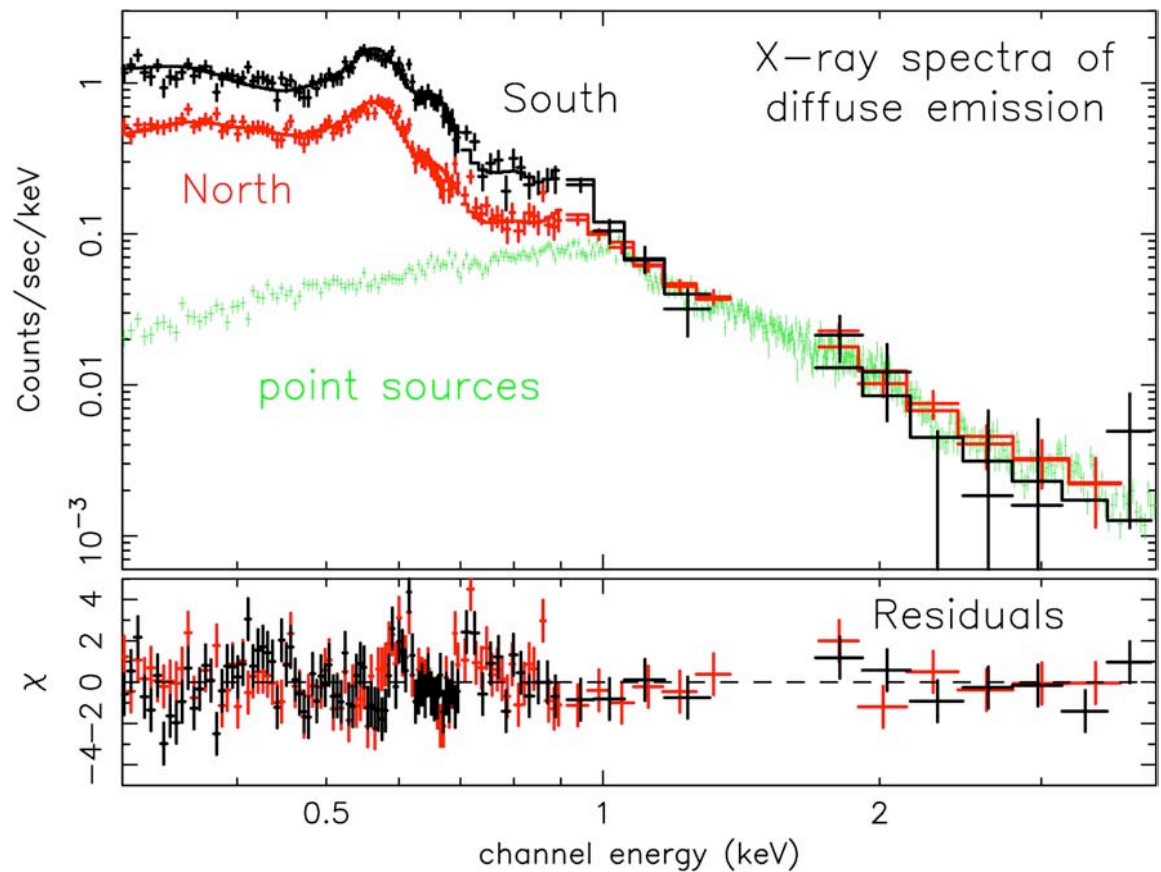

Figure 3: X-ray spectra of the diffuse emission. The lower and upper spectrum refers to, respectively, the northern, bright patch and the southern, more extended structure in Fig. 2. The error bars attached to each data point reflect 1σ errors from counting statistics. The solid histograms represent a model fit to the spectra, based on emission from a thermal plasma *(16)*. The lower panel gives the fit residuals. The green spectrum shows the contribution from stellar X-rays to the spectrum of the northern diffuse emission.

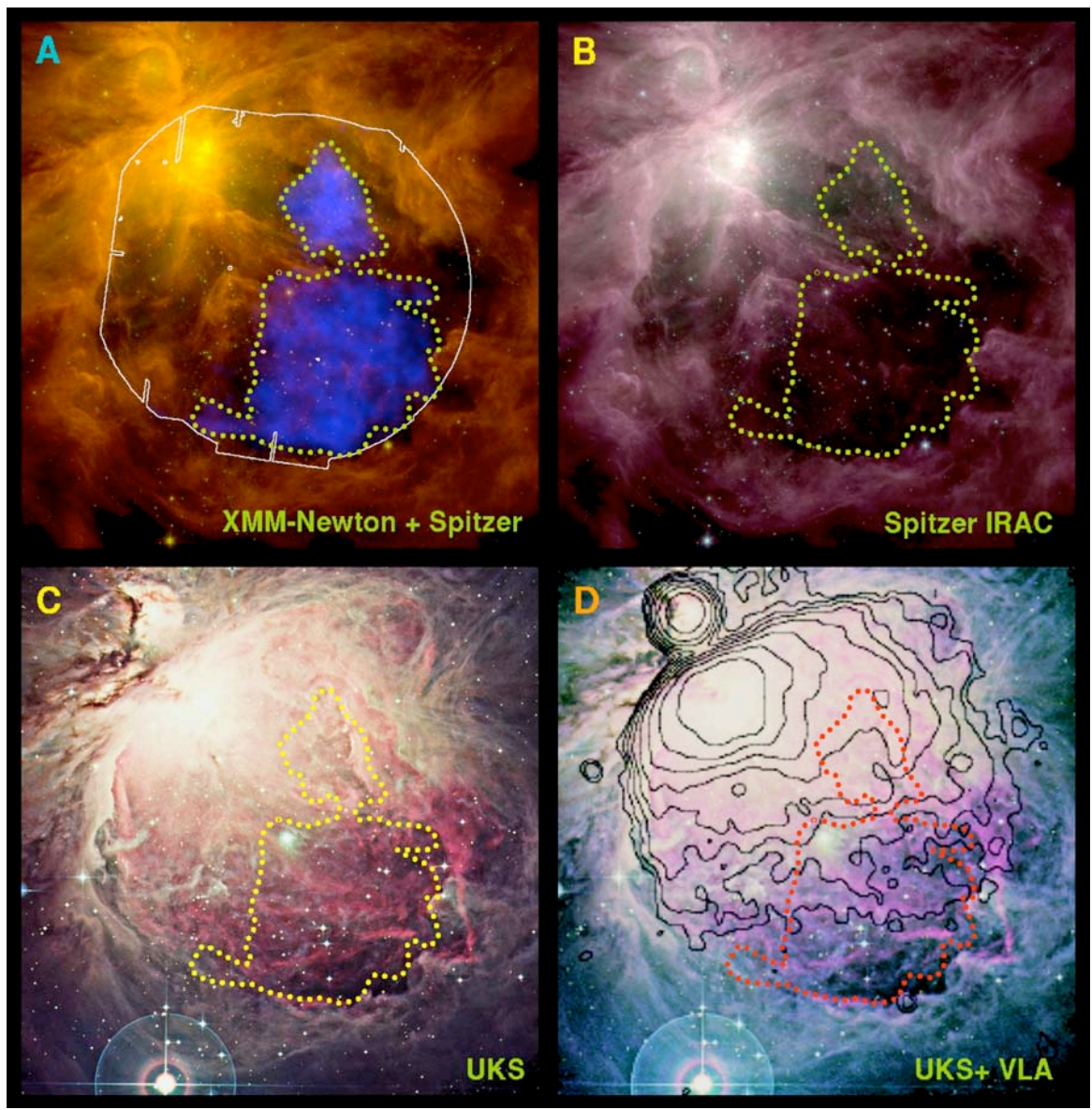

Figure 4: A multi-wavelength view of the Orion Nebula region. The four panels show identical regions; the width of each panel is about 42 arcmin, corresponding to 4.9 pc at a distance of 400 pc. Upper left (A): diffuse X-rays (in blue) superimposed on a Spitzer infrared image (Fig. 2B); upper right (B): An infrared image (3.6+4.5+5.8 μm composite from the Spitzer Space Telescope; lower left (C): An optical image (from the UK Schmidt telescope; copyright Anglo-Australian Observatory/David Malin Images); lower right (D): same with radio contour overlays (from the VLA, observed at 330 MHz *(9)*, reproduced by permission of the American Astronomical Society).

The dotted contour outlines the area where most of the detected diffuse soft X-ray emission is located.

**Supporting Online Material**

**Materials and Methods**

**1. Reduction of the XMM-Newton X-ray data**

All X-ray data discussed here were taken with the European Photon Imaging Cameras (EPIC) on board the X-Ray Multi-Mirror Satellite XMM-Newton, with an angular resolution of 5 arcseconds and a field of view of 30 arcminutes each. One of these Charge Coupled Device (CCD) cameras is of the PN type *(14)* and two are of the MOS type *(15)*. The middle field in Fig. 2A was observed on 2005 February 18 for ≈22 ks and on 2006 March 2 for ≈91 ks, the upper left field on 2001 October 3 for ≈39 ks, and the lower left field on 2001 September 15 for ≈22 ks.

Our data reduction involved standard procedures in the XMM-Newton Science Analysis System (SAS) software. Imaging analysis was performed using SAS v6.1. Spectral analysis took advantage of the improved treatment of extended sources offered in SAS v7.0.

Photon event lists in each of the three EPIC cameras were produced using the SAS EPCHAIN and EMCHAIN tasks for the PN and MOS data, respectively. Filtering of these event lists and exclusion of time intervals affected by high background were conducted as described by Güdel et al. *(S1)*.

The three-color images in Fig. 2A were constructed using data from eight XMM-Newton observations of four different, overlapping fields. In each of three energy bands – 0.3-1.0 keV, 1.0-2.0 keV and 2.0-7.3 keV – images were extracted from the event list for each camera and observation and mosaicked using the EMOSAIC task. Each image mosaic was adaptively smoothed *(S2)* to achieve a desired signal-to-noise ratio of 5 using ASMOOTH. A mask was employed to define areas of active CCD to be smoothed. Exposure maps account for the different exposure times of the observations (from 20 to 91 ks) and the lower sensitivity of the MOS cameras compared to the PN. These were produced for each camera, observation and energy band using EEXPMAP, and the MOS exposure times were converted to PN-equivalent exposure times by dividing by a band dependent factor (4.0, 2.9 and 2.7 for the 0.3-1.0 keV, 1.0-2.0 keV and 2.0-7.3 keV bands, respectively, as appropriate for the typical X-ray spectrum of young stars) before a mosaic was made for each band. The mosaicked exposure map was scaled to have a peak value of 1 to maintain counts as the units of the images. The exposure maps do not account for the decreasing sensitivity away from the detector center (the so-called vignetting) as this would distractingly accentuate the background component in the outer region of each observed field, because this particle-induced background is not subject to vignetting. The brightness of off-axis sources therefore appears lower (by up to a factor 2.5 at the very edge of the field of view) with respect to on-axis sources. Three-color images were generated by the Chandra CIAO software package task DMIMG2JPG *(S3)* using a square-root scaling and minimum and maximum pixel values of 0 and 5 in each energy band.

The X-ray image in Fig. 2B was produced using data from only the longest observation (67 ks useful PN exposure time; 91 ks MOS) of the field containing the diffuse soft X-ray emission. As described above, images in the 0.3-1.0 keV and 1.0-2.0 keV energy bands were smoothed using ASMOOTH, but this time a mask was employed that excluded regions around point sources from the smoothing and a signal-to-noise of 20 was desired, in order to show the diffuse emission. Point sources were detected and parameterized (actually in 0.5-2.0 keV, 2.0-7.3 keV and 0.5-7.3 keV images as part of a parallel investigation) using a combination of SAS tasks and FTOOLS described in Güdel et al. *(S1)*. Elliptical source exclusion regions were calculated using the REGION task (Fig. S1A). These defined the point spread function contour at which the surface brightness of counts from the source fell to 0.3 times that from the local background. These were made into a mask, using REGIONMASK, which was then combined with the mask defining the active CCD area.

The image in Fig. 2B shows the ratio of the 0.3-1.0 keV and 1.0-2.0 keV source-excluded smoothed images on a logarithmic intensity scale. This ratio is closely representative of the measured 0.3-1.0 keV flux because this excess emission is mostly seen in the softer band (Fig. S1B) while the harder band shows a nearly flat count distribution in the EON, dominated by background radiation (Fig. S1C). But the ratio image (Fig. 2B) has two key advantages over the 0.3-1.0 keV image alone (Fig. S2). Firstly, the soft-band image shows excess values not only where we see the soft diffuse emission in Fig. 2A but also around the confusion of sources close to Theta 1 Ori C. As the typical spectrum of the point sources (see e.g. Fig. 3) is harder than that of the diffuse emission (especially in this more heavily-absorbed region of the nebula), this excess, unlike that of the true soft diffuse emission, also shows up in the 1.0-2.0 keV image and is cancelled out in the ratio image. Secondly, the two single-band images each suffer from a similar amount of vignetting, so while the diffuse emission in the outer regions of the 0.3-1.0 keV image appears faint and is poorly seen, vignetting has a much smaller effect in the ratio image, enabling the diffuse emission close to the edge of the field of view to be much more clearly seen.

Somewhat coincidentally the typical value of the ratio in regions of normal background is approximately 1. The image in Fig. 2B shows a logarithmic scaling for the X-rays from 1.3 to 4, which shows the soft diffuse emission.

The extraction of X-ray spectra also used only this longest observation, and concentrated on data from the most sensitive EPIC instrument, the PN CCD camera. Polygonal extraction regions for the northern and southern areas of diffuse emission were defined by viewing the 0.3-1.0 keV image in the DS9 imaging software (Fig. S3). Regions around point sources were excluded as noted above. Nevertheless, due to the relatively large point-spread function of XMM-Newton and the detection of "out of time events" at false positions along the readout direction while the CCD is being read out, a considerable number of counts from the point sources still contaminate the spectra of the diffuse emission. This contamination was modeled by extracting a composite spectrum of all point sources in each of the northern and southern extraction regions and simultaneously fitting the diffuse and point-source spectra (see below). Circular extraction regions with the small radius of 10 arcsec, which enclose approximately half of the total counts from each source, were used to minimize the contribution of diffuse emission to the point-source spectra.

A background spectrum was taken from a region where no soft diffuse emission was visible in a largely source-free area at the upper right of the detector (Fig. S3). As this region is further off-axis than the extraction regions for the diffuse emission and some components of the background are vignetted, the background was spectrum was scaled by factors 1.18 and 1.16 for the northern and southern diffuse regions, respectively. This scaling assigns all counts in the diffuse spectra in the 4-7 keV region to the background (i.e. it is the maximum possible correction), which is consistent with the expected vignetting in this energy range and off-axis angle *(S4)*. The background spectrum was scaled to the areas of the two diffuse-source areas and was then subtracted.

All PN spectra were extracted by selecting only CCD events with event patterns 0-4 (attributed to source X-rays rather than cosmic ray hits of the CCD) while MOS spectra required no further filtering.

The spectral response of the instrument was modeled for each spectrum by generating response matrices using the RMFGEN task and ancillary response files using ARFGEN, with the spatial variation over the extraction area modeled as a flat detector map with 20 x 20 pixels.

## 2. Reduction of the Spitzer Space Telescope infrared data

The Spitzer observations of the Orion nebula were obtained as part of guaranteed time observations (program 43, PI Fazio) in late 2004 and are described further in *(S5)*. Standard pipeline products processed by the Spitzer Science Center (SSC) are basic calibrated data (BCD) products which include flux-calibrated individual frames, errors, etc., and so-called "Post-BCD" (Post-basic calibrated data) products, which include mosaics of individual frames, mosaics of errors, etc. The pipeline has applied all of the standard corrections astronomers usually use when creating mosaics, for example, applying outlier rejection to remove the signatures of cosmic rays, and producing flux-calibrated images. We downloaded post-BCD from the Spitzer Science Center pipeline version 14.0. We used the SSC mosaicking and point-source extraction (MOPEX) software *(S6)* to combine the data into large images, separately for the different bands provided by the IRAC instrument (3.6μm, 4.5μm, 5.8μm, and 8μm). The resultant integration time is 24 seconds per position and per channel.

## 3. Spectral fits for the X-ray data

The X-ray spectra were binned in energy, differently at different energies (in the 0.25-0.7 keV range: minimum of 30 counts per bin before background subtraction; in the

0.7-0.9 keV range: minimum of 100 counts for northern source and 500 counts for southern source; in the 0.9-1.38 keV and 1.58-4.0 keV ranges: minimum of 500 counts for the northern source, 2000 counts for the southern source; see Fig. 3). We have removed the bins around 1.5 keV due to strong contamination by an instrumental line of aluminum. The error bars attached to each data point reflect 1$\sigma$ errors from counting statistics. The synthesized source spectra of the combined point sources were rebinned to a minimum number of 30 cts per bin.

Independent spectral fits showed that the diffuse emission originates from a dominant cool (1-2 MK) plasma but shows contributions from hotter plasma that are similar to the emission in the stellar point sources. As suggested above, this is expected because the extended instrumental wings of the point sources contaminate the diffuse areas. On the other hand, some diffuse emission is also included in the extracted stellar point-source spectra. We therefore modeled the spectra together, allowing for a variable contribution of the stellar emission to the spectra of the diffuse emission and vice versa, independently for the northern and southern areas.

Spectral fits were made in the XSPEC software *(S7)* using the APEC atomic emission line code for a hot, thermal plasma in collisional ionization equilibrium included in the software package. The diffuse emission was defined by one isothermal component (for each of the areas), while the stellar spectra required three isothermal components, each component described by its electron temperature and emission measure. An absorbing interstellar hydrogen column density, $N_H$, was simultaneously fitted, again independently for the diffuse emission and the point sources. Finally, we also considered element abundances. Most elements do not generate strong emission lines in our soft spectra of the diffuse emission. We therefore adopted their abundances as determined previously for the hot, X-ray emitting wind of $\theta^1$ Ori C *(S8)* with respect to the solar photospheric fraction *(S9)*, namely Ne (1.04 times the solar fraction), Mg (0.94), Si (1.11), S (1.22), Ar (1.48), Ca (1.72), and Fe (0.62). We fitted C, N, and O abundances to the diffuse emission spectrum because these elements produce the dominant spectral contributions below 0.7 keV. For the stellar spectrum, we additionally fitted abundances of Ne, Mg, S, Si, Ar, and Fe that produce line emission features in the harder portion of the spectra. The remaining stellar element abundances were fixed at values commonly seen in young stellar X-ray sources *(S1)*. For the stellar spectra, we also held C and N fixed at such values because the dominant hot stellar plasma produces no discernible features of C and N at the soft end of the spectrum.

The results are given in Table S1. The errors give 1$\sigma$ errors due to counting statistics and represent to the formal uncertainties of the parameters in the adopted, simplistic model. Unknown systematic uncertainties in the atomic emission line model and in the detector calibration would increase the error bars. The X-ray luminosity was determined by integrating the unabsorbed best-fit model spectrum over the energy range of 0.1-10 keV, assuming a distance of 400 pc *(S10-S12)*.

The stellar point-source spectra, not described in the Table, required three components with temperatures of 2.2 MK, 8.7 MK, and 27 MK with a ratio of emission measures of 1:3.4:9.4 for the northern area, and temperatures of 4.9 MK, 9.4 MK, and 30 MK with a ratio of emission measures of 1:1.4:3.1 for the southern area. The hottest components clearly dominate both stellar spectra, and the spectral flux is monotonically decreasing with decreasing photon energy below 1keV, the spectra thus showing a pronounced peak at 1 keV due to emission lines of iron and neon (Fig. 3). The stellar hydrogen absorption column densities amount to $1.1 \times 10^{21}$ cm$^{-2}$ and $8.5 \times 10^{20}$ cm$^{-2}$ for the N and S stellar sources, some of this material normally being confined to the immediate circumstellar environment. The coolest component of the stellar point source spectrum contributes ≈3% and ≈4% to the emission measure of the cool component collected from the diffuse areas. The three components of the scaled stellar point source spectrum together contribute only about 5% to the soft spectrum of the diffuse emission at 0.3-0.6 keV while they accurately explain all of the hard emission above 1 keV (Fig. 3), corroborating our view that the hard diffuse emission is due to contamination from stellar point sources.

Table S1 also lists some further parameters used in conjunction with the X-ray results to estimate physical source parameters. For these parameters (projected area, estimated volume, electron density, and plasma mass), we give values referring to a slab model (EON circular area with a radius of 2 pc and a depth of 0.9 pc, first number) and for a spherical model (a spherical volume filling the EON with a radius of 2 pc, second number).

**Supporting figures**

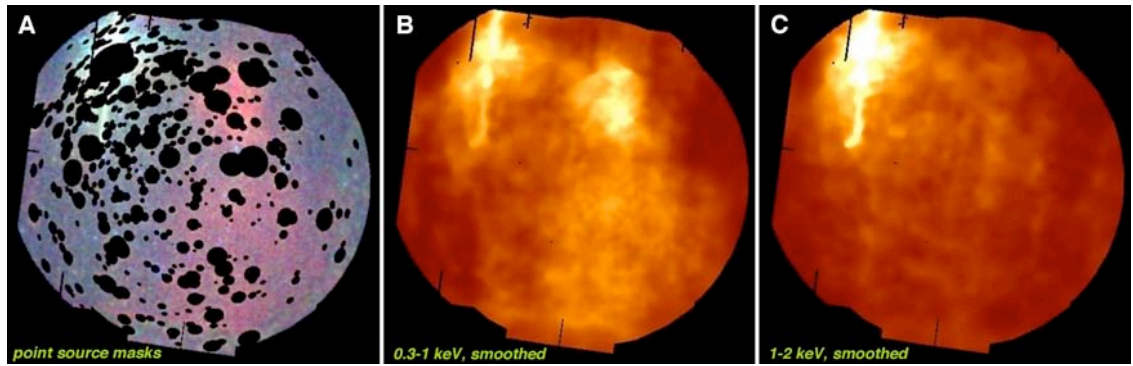

Figure S1: Left (A): X-ray image after excising the areas around the point sources. Middle (B) and Right (C): Smoothed X-ray images of the EON region. Both figures have been adaptively smoothed after removal of X-ray point sources. Panel (B) is for the 0.3-1.0 keV band, showing soft diffuse emission (in the right half) but also strong emission from the Trapezium cluster stars. Panel (C) shows the 1-2 keV band in which the Trapezium stars dominate.

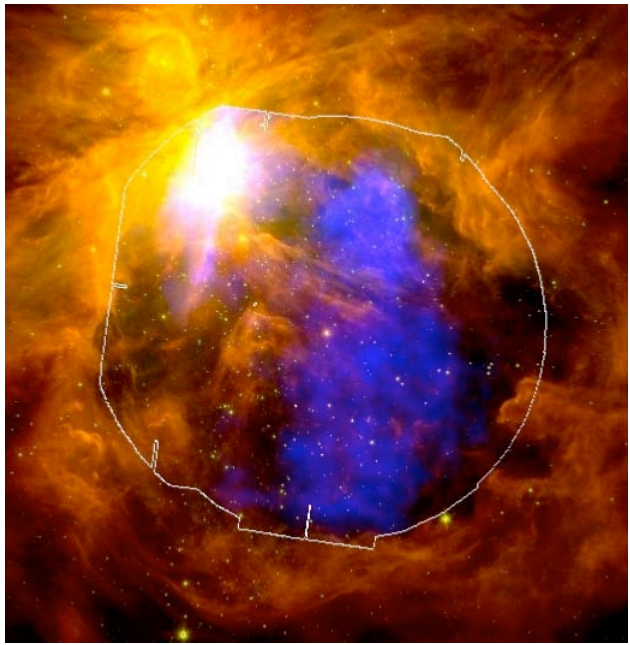

Figure S2: X-ray diffuse emission (blue) superimposed on a Spitzer composite image. This figure is analogous to Fig. 2B except that the flux in the 0.3-1 keV X-ray band is shown rather than its excess relative to the 1-2 keV band. The lack of vignetting correction suppresses X-rays near the border of the detector, and the broadband but predominantly harder X-rays from the Trapezium cluster saturate the upper left area.

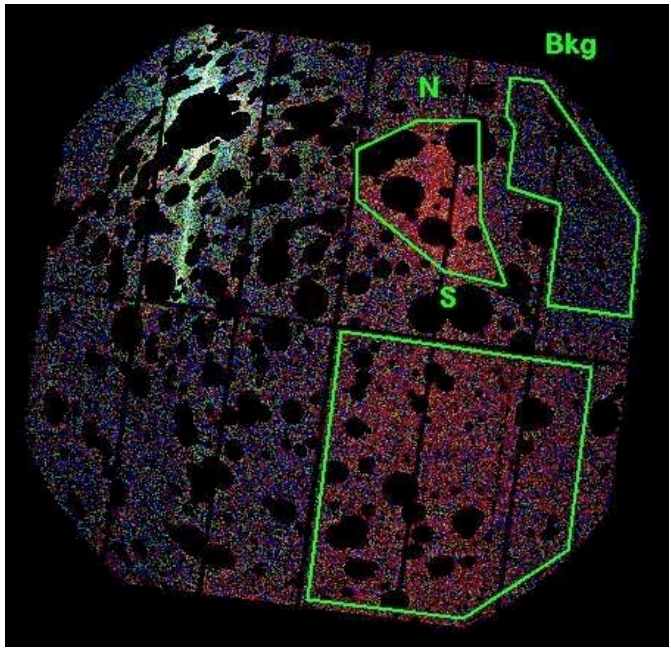

Figure S3: Extraction regions on the PN detector used for spectral analysis. The green polygons show the extraction regions for the northern diffuse source (upper left), the southern diffuse source (lower area), and the background area (upper right). The point sources have been removed, and the remaining counts on the detector have been color-coded as in Fig. 2A.

**Supporting table**

Table S1: Spectral interpretation and source geometry. The first seven parameters have been derived from fits of the observed X-ray spectra of diffuse emission with calculated model spectra of hot plasmas. The last four parameters are based on the observed and adopted geometric extent of the extended X-ray sources. Equal signs in the column for the southern area indicate identical values as for the northern area. Two numbers separated by a slash refer to two different geometric model assumptions (see text).

| Parameter | | Unit | Northern Area | Southern Area |
|---|---|---|---|---|
| Hydrogen column density | $N_H$ | $10^{20}$ cm$^{-2}$ | 4.1(±0.7) | 0.4(-0.4,+0.5) |
| Volume emission measure | EM | $10^{54}$ cm$^{-3}$ | 1.45(±0.25) | 1.88(±0.28) |
| Electron temperature | T | $10^6$ K | 1.73(±0.03) | 2.08(-0.02,+0.03) |
| Oxygen abundance | | - | 0.43(-0.04,+0.06) | = |
| Nitrogen abundance | | - | 1.5(-0.2,+0.3) | = |
| Carbon abundance | | - | 1.5(-0.2,+0.3) | = |
| X-ray luminosity | $L_X$ | $10^{31}$ erg s$^{-1}$ | 2.3 | 3.2 |
| Cross-sectional area | A | pc$^2$ | 0.24 | 1.38 |
| Volume | V | pc$^3$ | 0.22/0.97 | 1.24/5.5 |
| Electron density | $n_e$ | cm$^{-3}$ | 0.47/0.22 | 0.23/0.11 |
| Observed plasma mass | M | $M_\odot$ | 0.003/0.006 | 0.008/0.017 |

**Supporting References and Notes:**